\documentclass[fleqn,usenatbib]{mnras}

\usepackage{newtxtext,newtxmath}

\usepackage[T1]{fontenc}

\DeclareRobustCommand{\VAN}[3]{#2}
\let\VANthebibliography\thebibliography
\def\thebibliography{\DeclareRobustCommand{\VAN}[3]{##3}\VANthebibliography}

\usepackage{graphicx}	
\usepackage{amsmath}	
\usepackage{subcaption}
\defcitealias{suazo24}{S24}

\title[Project Hephaistos – III.]{Project Hephaistos – III. Characterizing anomalous infrared sources identified as Dyson-sphere candidates}

\author[Korn et al.]{Andreas J. Korn$^{1}$\thanks{E-mail: andreas.korn@physics.uu.se},
Matías Suazo$^{1}$,
Erik Zackrisson$^{1}$,
Pía Cortés-Zuleta$^{2}$,
Adam D. Rains$^{1}$,
\newauthor
Armin Nabizadeh$^{1}$
\\
$^{1}$Observational Astrophysics, Department of Physics and Astronomy, Uppsala University, Box 524, SE-751 20 Uppsala, Sweden\\
$^{2}$SUPA, School of Physics and Astronomy, University of St. Andrews KY16 9SS, UK.\\
}
\date{Accepted XXX. Received YYY; in original form ZZZ}

\pubyear{2015}

\begin{document}
\label{firstpage}
\pagerange{\pageref{firstpage}--\pageref{lastpage}}
\maketitle

\begin{abstract}
The infrared-flux excess of stars harbouring Dyson spheres represent one potential technosignature of extraterrestrial intelligence. In a previous Project Hephaistos paper, we highlighted seven M dwarfs within 300 pc with unusual infrared properties that seem to resemble those expected for Dyson spheres. In the present study, we present an analysis of photometric and, in some cases, spectroscopic data on these seven objects, plus three additional objects with similar properties, to further constrain their nature. The stellar parameters, derived from calibrated empirical relationships for M dwarfs, reveal no irregularities with respect to the overall M-dwarf population of main-sequence stars. While the infrared properties of our targets resemble those of circumstellar disks in a transitional state, the objects with spectroscopic data show no signs of youth usually associated  with such objects. One of our spectroscopic targets does exhibit weak H$\alpha$ emission, but this is more likely attributed to stellar activity than to the gaseous accretion disk expected for a young star. After this analysis, we still find no clear explanation for the infrared excess of these stars, but note that observations with the JWST and/or the Atacama Large Millimeter/submillimeter Array would be able to probe scenarios in which the infrared excess is due to circumstellar dust emission or an infrared-bright background source that remains undetected at shorter wavelengths. For two of our stars, JWST observations have recently revealed superpositions with very red background galaxies. 
\end{abstract}




\section{Introduction}
The strive to detect life beyond Earth can generally be divided into the search for signs of biological activity (biosignatures) and signs of the technology developed by an intelligent species (technosignatures; the Search for Extra-Terrestrial Intelligence, SETI). In terms of detection prospects, it remains unclear which strategy is the more likely to succeed. While it is usually assumed that life starts off in a primitive biological state and gradually -- and possibly only very rarely -- reaches a state where a species develops technology, technosignatures can in principle last long and be detected over much larger distances than biosignatures \citep[e.g.][]{Wright22}. 
 
Dyson spheres \citep{Dyson60} are hypothetical megastructures constructed around stars to harvest radiation energy \citep[for a review, see][]{Wright20}, possibly for the purpose of powering habitats, vessel propulsion, computing or transmitting long-range signals. The primary technosignature of such structures is the waste heat emitted, usually assumed to be detectable as excess continuum radiation in the infrared (IR) part of the electromagnetic spectrum. While waste heat rates favorably on six of the nine axes of merits for technosignatures described by \citet{Sheikh20}, such signatures are ambiguous since there are many astrophysical phenomena that may produce similar effects. 

As part of {\it Project Hephaistos\footnote{\url{https://www.astro.uu.se/~ez/hephaistos/hephaistos.html}}}, \citet[hereafter \citetalias{suazo24}]{suazo24}  conducted a search for Dyson spheres, focusing on the anticipated infrared excess of such objects, and identified seven stars that exhibited anomalous infrared emission compatible with the theoretical spectral energy distribution of Dyson spheres described by \citetalias{suazo24}. These excesses were found in the mid-infrared range (specifically at 12 and 22 microns) using data from AllWISE, an extension of the IR-satellite WISE program \citep{wright_wise} that combines data from different phases of the mission. The current paper aims to offer further insights into these stars and constrain potential mechanisms responsible for their observed infrared excess.

An excess in the infrared regime is commonly associated with circumstellar dust around a star, indicative of stars in the protoplanetary stage or a later phase known as a debris disk \citep[e.g.,][]{cotten2016}. The final stage of a debris disk comprises second-generation dust and little gas as accretion onto the star has ceased. This infrared excess is characterized by the fractional luminosity ($f = L_{\rm IR}/L_{\star}$), providing insights into the star's evolutionary stage. Typically, a fractional luminosity $f \sim 0.01$ serves as the threshold to differentiate between early and later phases \citep{wyatt2008,hughes2018}. The analysis by \citetalias{suazo24} indicates that these stars are unlikely to be in the protoplanetary phase, based on the various criteria in the down-selection pipeline, including stellar variability and H$_{\alpha}$ emission when available. In the context of astrophysical explanations, this supports the hypothesis that these stars host debris disks with exceptionally high $L_{\rm IR}/L_{\star}$ ratios (so-called extreme debris disks, EDDs, \citet{balog09}). However, while EDDs are found around stars with spectral types from G to K, a notable feature of the \citetalias{suazo24} study is that all its candidates are M-dwarf stars, identified based on their position in the Gaia colour-magnitude diagrams and stellar parameter estimates from Gaia DR3 \citep{gaia_space,gaiadr3,gaiadr3_apsis1, gaiadr3_apsis2}.

Infrared-excess detection rates are well-established for solar-type stars \citep{bryden2006,trilling2008} and A-type stars \citep{su2006}. However, due to the low luminosity of M dwarfs, such rates have not been precisely determined for this stellar type \citep{lestrade2006}, with constraints only falling within the range of 0 to 14\% \citep{lestrade2006,avenhaus2012,kennedy2018}. To date, only a limited number of confirmations exist, primarily in the submillimeter regime \citep[e.g.,][]{luppe20,cronin2022,cronin2023}. Various factors contribute to the challenge of detecting debris disks around M dwarfs, including potential detection biases \citep{heng13,kennedy2018} and age biases \citep{riaz06,avenhaus2012}. Moreover, studies have suggested that the physical processes governing debris disk evolution around M dwarfs may significantly differ from those observed in their more massive counterparts \citep{plavchan2005}.

In Section~\ref{sec:observations}, we describe the observational data (including new observations) analyzed for our ten Dyson-sphere candidates. In Section~\ref{sec:results}, we characterize these sources in terms of stellar parameters, optical spectral features and infrared colors. In Section~\ref{sec:discussion}, we discuss these results and explain how future follow-up observations may help us disentangle their true nature. Section~\ref{sec:conclusions} summarizes our results.

\section{Observational data}
\label{sec:observations}

This paper examines the same candidates as presented in \citetalias{suazo24}, summarized in Table~\ref{tab:stars}, which includes the temperature of the Dyson sphere $T_{\rm DS}$ and the covering factor $\gamma = L_{\rm DS}/L_{\star}$ that best fit their Dyson sphere models. In addition to the original table in \citetalias{suazo24}, this table now includes Transiting Exoplanet Survey Satellite (TESS) Input Catalogue IDs, as these stars are part of the TESS survey focused on detecting exoplanets through optical flux drops \citep{tess,stassun2018,stassun2019}. We organize our analysis into several tasks. Firstly, we re-derive the stellar parameters for these stars utilizing available data from Gaia Data Release 3 \citep{gaia_space,gaiadr3}, 2MASS \citep{2mass},
and AllWISE \citep{cutri}. Given that all Dyson-sphere (DS) candidates turned out to be M dwarfs, we restrict this re-derivation to empirical relations specifically tailored for this spectral type (Section~\ref{sec:photometry}).

\begin{table*}
\caption{Dyson-sphere candidates from \citet{suazo24}. Data derived from
    $\rm ^{a}$ Gaia EDR3 \citep{Bailer-Jones21}, $\rm ^{b}$ Gaia DR3 $\rm ^{c}$ \citet{suazo24} and\linebreak $\rm ^{d}$ AllWISE \citet{cutri}. $T_{\rm DS}$ and $\gamma$ correspond to the temperature of the Dyson sphere and the covering factor ($L_{\rm DS}/L_{\star}$) that best fits the photo\-metric data.}
    \centering
    \begin{tabular}{c|c|c|c|c|c|c|c|c|c|c}
        \hline
        Label & Gaia DR3 ID & Distance$\rm ^{a}$ [pc] & $m_G$$\rm ^{b}$ & $T\rm _{eff}$$\rm ^{b}$ [K] & $T_{\rm DS}$$\rm ^{c}$ [K] & $\gamma$$\rm ^{c}$ & SNR$\rm ^{d}$ (W3/W4) & TIC ID \\
        \hline
        A & 3496509309189181184 & 142.9 $\pm$ 1.0 & 15.99 & - &  138 $\pm$ 6 &  0.080 $\pm$ 0.009 & 22.5 / 16.6 & 9348687 \\
        B & 4843191593270342656 & 211.6 $\pm$ 3.5 & 17.71 & 3574 &  275 $\pm$ 40 & 0.068 $\pm$ 0.008 & 13.9 / 3.8 & 140206070 \\ 
        C & 4649396037451459712 & 219.4 $\pm$ 6.2 & 18.39 & 3238 & 187 $\pm$ 16 & 0.143 $\pm$ 0.016  & 10.5 / 5.0 & 140754886 \\
        D & 2660349163149053824 & 211.5 $\pm$ 5.8 & 17.66 & 3473 & 178 $\pm$ 20 & 0.163 $\pm$ 0.028 & 10.4 / 4.8 & 422524059 \\
        E & 3190232820489766656 & 274.7 $\pm$ 6.1  & 17.00 & 3556 & 180 $\pm$ 26 & 0.081 $\pm$ 0.020 & 10.3 / 3.6 & 34163716 \\
        F & 2956570141274256512 & 265.0 $\pm$ 2.6 & 16.32 & 3674 & 176 $\pm$ 21 & 0.167 $\pm$ 0.032 & 5.7 / 4.5 & 13320963 \\
        G & 2644370304260053376 & 249.9 $\pm$ 3.7 & 16.48 & 3480 & 100 $\pm$ 9 & 0.131 $\pm$ 0.028 & 5.0 / 3.5 & 398584776 \\
        \hline
    \end{tabular}
    \label{tab:stars}
\end{table*}

Additionally, we obtained long-slit spectra using the Alhambra Faint Object Spectrograph and Camera (ALFOSC) mounted on the Cassegrain focus of the 2.5-m Nordic Optical Telescope (NOT) at the Observatorio del Roque de los Muchachos on La Palma, Spain. We took low-resolution spectra ($R \sim$ 2000) for the candidates A and D using grism \#8 that covers the range between 5690 and 8580 $\rm \Angstrom$. Given the apparent faintness of our candidates ($G$ $> 16$), we opted for this low-resolution mode. Despite the faintness, the spectra exhibit acceptable signal-to-noise ratios (SNR\,$\sim$\,30).

We reduced the spectra taken with ALFOSC using the PyNOT package\footnote{https://github.com/jkrogager/PyNOT}. This package performs standard reduction steps, including overscan correction, bias subtraction and division by a normalized flat field. Wavelength calibration was achieved using HeNe-lamp calibration frames which establish the dispersion solution. Although PyNOT is optimized for a wider wavelength range, we fine-tuned the pipeline parameters to optimize its performance within our spectral range, considering the low signal-to-noise ratio. Additionally, PyNOT incorporates astroscrappy \citep{astroscrappy}, a Python package for cosmic ray detection based on the algorithm by \citet{vandokkum2001}. Further details of this analysis are discussed in Section~\ref{sec:spec}. 

\begin{figure}
    \centering
    \includegraphics[width=\columnwidth]{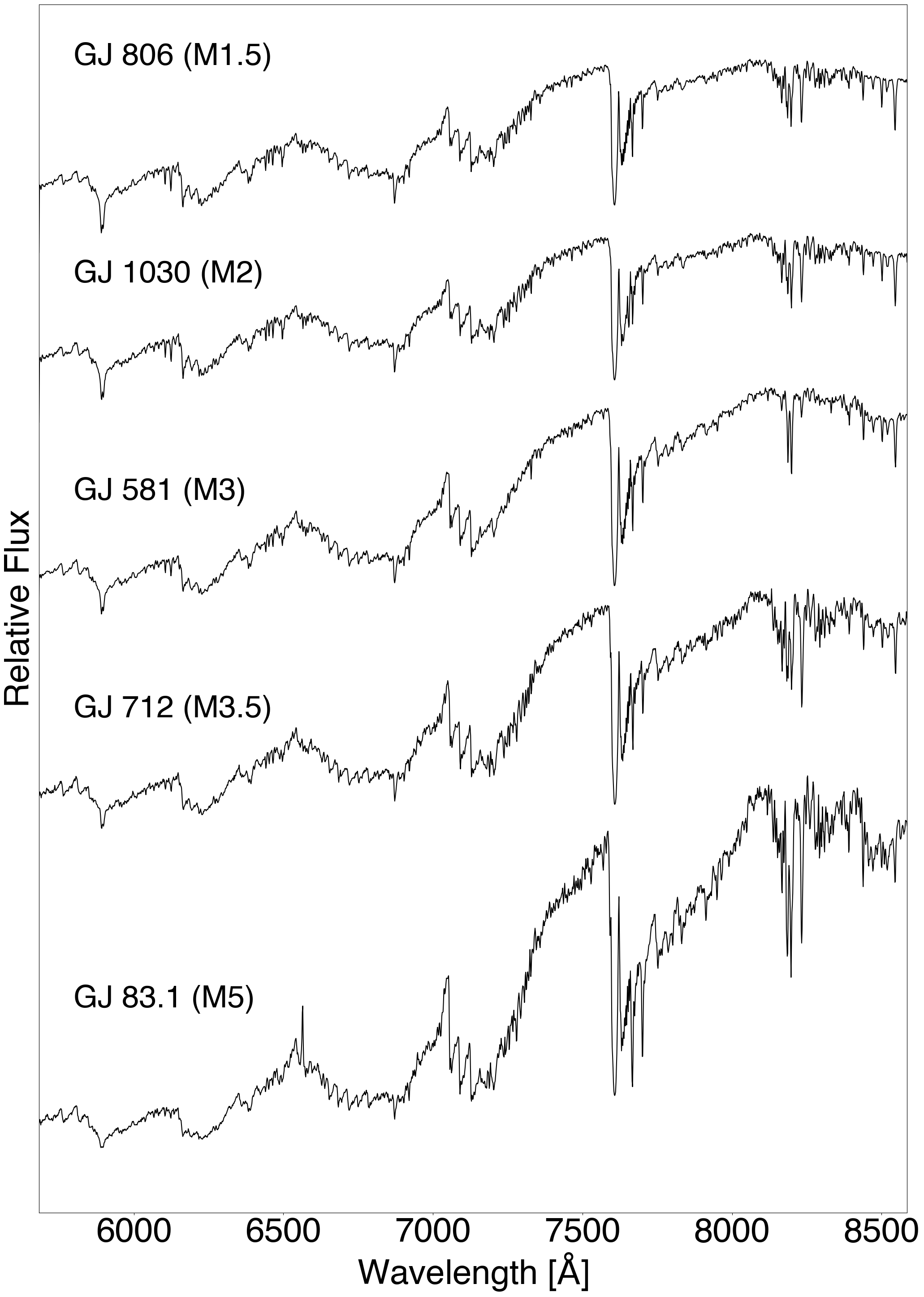}
    \caption{Stellar spectra of the five standard M dwarfs taken with ALFOSC.}
    \label{fig:mspectra}
\end{figure}

To establish a comparison metric, we also obtained spectra with the same instrument for five standard M-dwarf stars spanning spectral subclasses 1.5 to 5.0. Table~\ref{tab:standard} provides a summary of information about these stars, along with their respective references. Notably, GJ 581 is among these stars and is one of the few M dwarfs with a confirmed debris disk \citep{Lestrade2012}. The debris disk around GJ 581 was spatially resolved using deep images at 70, 100, and 160 $\mu m$ obtained with the Photodetector and Array Camera and Spectrometer (PACS) on the Herschel Space Observatory \citep{poglitsch}. \citet{Lestrade2012} derived that the debris disk spans from 25 ($\pm 12$) to 60 AU, with a temperature range from 30 to 50 K and a fractional luminosity $f \sim 10^{-4}$. This combination of temperature and fractional luminosity does not produce detectable infrared excess in WISE observations. Figure~\ref{fig:mspectra} displays the spectra of the five standard stars from Table~\ref{tab:standard}, all acquired with the same ALFOSC settings as our targets.

\begin{table*}
\caption{Standard M dwarf stars. The signal-to-noise ratio (SNR) corresponds to that of our spectra. Data sources: $^{\rm a}$ Gaia EDR3 \citep{Bailer-Jones21},\linebreak $^{\rm b}$ Gaia DR3, $^{\rm c}$ \citet{gaidos2014_all}, $^{\rm d}$ \citet{mann2019}, $^{\rm e}$ \citet{newton2014} and $^{\rm f}$ \citet{lepine2013}.}
    \centering
    \begin{tabular}{l|c|c|c|c|c|c|c|c}
        \hline
        Gliese ID & Distance$\rm ^{a}$ [pc] & $m_G$$\rm ^{b}$ & Mass [M$_{\odot}$]$ ^{\rm c}$ & $T\rm _{eff}$ [K] $^{\rm c}$ & Radius [R$_{\odot}$] $^{\rm c}$ & [Fe/H] & Spectral Type & SNR \\
        \hline
        
        GJ 806 & 12.06 & 9.83 & 0.432 $\pm$ 0.043$^{\rm d}$ & 3542 $\pm$ 61$^{\rm d}$ & 0.443 $\pm$ 0.018$^{\rm d}$ &  $-$0.15 $\pm$ 0.08$^{\rm d}$ & M1.5$^{\rm f}$/M2$^{\rm c}$ & 171 \\
        
        GJ 1030 & 22.23 & 10.51 & 0.57 $\pm$ 0.07 & 3807 $\pm$ 78&  0.53 $\pm$ 0.04 & 0.07 $\pm$ 0.12 $^{\rm c}$ & M2$^{\rm c}$ & 120 \\ 
        
        GJ 581 & 6.30 & 9.42 & 0.32 $\pm$ 0.06 & 3413 $\pm$ 61 & 0.33 $\pm$ 0.05 & $-$0.21 $\pm$ 0.11$^{\rm c}$ & M3$^{\rm c}$ & 141 \\
        
        GJ 712 & 14.80 & 11.38 & 0.19 $\pm$ 0.12 & 3281 $\pm$ 101 & 0.22 $\pm$ 0.09 & $-$0.48 $\pm$ 0.15$^{\rm e}$ & M3.5$^{\rm f}$/M4$^{\rm e}$ & 64 \\
        
        GJ 83.1 & 4.46 & 10.68 & 0.14 $\pm$ -- & 3074 $\pm$ 60 & 0.19 $\pm$ -- & $-$0.16 $\pm$ 0.08$^{\rm d}$ & M5$^{\rm e,f}$ & 80 \\
        \hline
    \end{tabular}
    \label{tab:standard}
\end{table*}


In addition to the seven candidates proposed by \citetalias{suazo24}, we incorporated three stars (H, I and J) into our ana\-lysis that did not meet the WISE-observations SNR criterion in the down-selection pipeline. These stars are included in our analysis because we were able to acquired ALFOSC spectroscopic data for these objects at little extra cost of observing time. Table~\ref{tab:bad} provides information about these three stars, including the predicted Dyson-sphere temperature and covering factor from the grid search implemented in \citetalias{suazo24}.
 
\begin{table*}
    \caption{Additional stars with available spectroscopic data. Data derived from
    $\rm ^{a}$ Gaia EDR3 \citet{Bailer-Jones21}, $\rm ^{b}$ Gaia DR3, $\rm ^{c}$ Suazo et al. 2024 and\linebreak
    $\rm ^{d}$ AllWISE \citet{cutri}. }
    \centering
    \begin{tabular}{c|c|c|c|c|c|c|c|c|c|c}
        \hline
        Label & Gaia DR3 ID & Distance$\rm ^{a}$ [pc] & $m_G$$\rm ^{b}$ & $T\rm _{eff}$$\rm ^{b}$ [K] & $T_{\rm DS}$$\rm ^{c}$ [K] & $\gamma$$\rm ^{c}$ & SNR$\rm ^{d}$ (W3/W4) & TIC ID \\
        \hline
        H & 2437221214075471744 & 270.3 $\pm$ 7.5 & 17.31 & 3418 &  130 $\pm$ 21 &  0.103 $\pm$ 0.027 & 2.4 / 3.3 & 20494175 \\
        I & 3854090071297359616 & 169.2 $\pm$ 2.6 & 17.40 & 3327 &  99 $\pm$ 20 & 0.147 $\pm$ 0.046 & 2.4 / 3.3 & 469015958 \\ 
        J & 651765552072217216 & 288.2 $\pm$ 4.0 & 16.13 & 3677 & 114 $\pm$ 28 & 0.058 $\pm$ 0.048  & 2.2 / 2.8 & 20494175 \\
        \hline
    \end{tabular}
    \label{tab:bad}
\end{table*}

\section{Results}
\label{sec:results}

\subsection{Stellar Parameters}
\label{sec:photometry}

To improve the understanding of our targets, we reassessed the stellar parameters using methods tailored for small and cool M-dwarf stars, consistent with those in this study. Any deviations from typical parameters for these stars could prompt a reevaluation of our initial classification. This investigation also serves to unveil alternative explanations for the observed anomalous infrared light seemingly associated with these stars.

In recent times, M dwarfs have garnered considerable attention in various exoplanetary campaigns. However, unraveling the properties of exoplanets requires the precise determination of the stellar characteristics of their host stars. Estimating planet properties is intricately linked to both the features of the host stars and key observables like the transit depth or Doppler radial-velocity amplitude. This reinforces the significance of obtaining accurate stellar parameters for exoplanetary campaigns. Consequently, establishing empirical relationships between observables and stellar parameters has emerged as an essential and valuable pursuit for characterizing planets orbiting the most common type of star in the Galaxy.

Furthermore, stellar parameters are also useful in gaining insights into the origin and diversity of planetary systems. For instance, the occurrence of giant planets appears to correlate with stellar mass \citep[e.g.,][]{johnson2010, gaidos2013}, while smaller, close-in planets become more prevalent with decreasing stellar mass \citep{howard2012}. To refine the stellar parameters of our candidates, we utilize various empirical relations. 

\subsubsection{Mass}
\label{sec:mass}

Determining masses for M-dwarf stars is important, especially in calculating planet masses through radial-velocity surveys. While mass estimates for individual stars can be obtained from stellar evolution models \citep[e.g.,][]{muirhead2012}, discrepancies observed between empirically derived and model-predicted mass-radius as well as radius–luminosity relationships for late-type stars \citep[e.g.,][]{boyajian2012, feiden2012} generates concerns about the reliability of mass determinations based on models. A dependable empirical alternative for estimating masses of individual stars involves establishing a relationship between mass and luminosity \citep[e.g.,][]{henry1993, delfosse2000}, which is calibrated using precise dynamical mass measurements obtained from binary-star systems. 

Following this line of thought, we determine the masses of our candidates using data from 2MASS. We employed the M\_-M\_K- code\footnote{\url{https://github.com/awmann/M\_-M\_K-?tab=readme-ov-file\#m\_-m\_k-}} which employs the M$_{K_{\rm s}}$ - M${_\star}$ relation established by \citet{mann2019}. This relationship tells us the mass of a star based on how bright it is in the $K\rm _s$ band. The brightness in this band is helpful because it is not affected by the metal content in the star \citep{mann2019}.

Table~\ref{tab:stellar} summarizes all the stellar parameters derived from photometric relations, including the masses we inferred. We compared masses with and without considering the metal content in our calculations from the iron abundances derived in Section~\ref{sec:metal}, and as we expected, it did not have any significant impact on our mass estimates. Table~\ref{tab:stellar} incorporates only mass estimates without considering the metallicity contribution since its precise value does not play a significant role (see above). 

\subsubsection{Radius}

Like the critical precision required for determining exoplanet masses, accurately determining stellar radii is pivotal in estimating planet sizes through transit methods. In pursuing stellar radii, we employ the empirical relation introduced by \citet{mann2015}. Analogous to our approach in determining stellar masses (Section~\ref{sec:mass}), this empirical relation primarily relies on the absolute magnitude of the star in the $K_{\rm s}$ band, exhibiting minimal sensitivity to the metallicity of the star. Consequently, considerations of metallicity are omitted in the estimation of this stellar parameter. The resulting values for stellar radii are detailed in Table~\ref{tab:stellar}.


\subsubsection{Surface Gravity}

\begin{figure}
    \centering
    \includegraphics[width=\columnwidth]{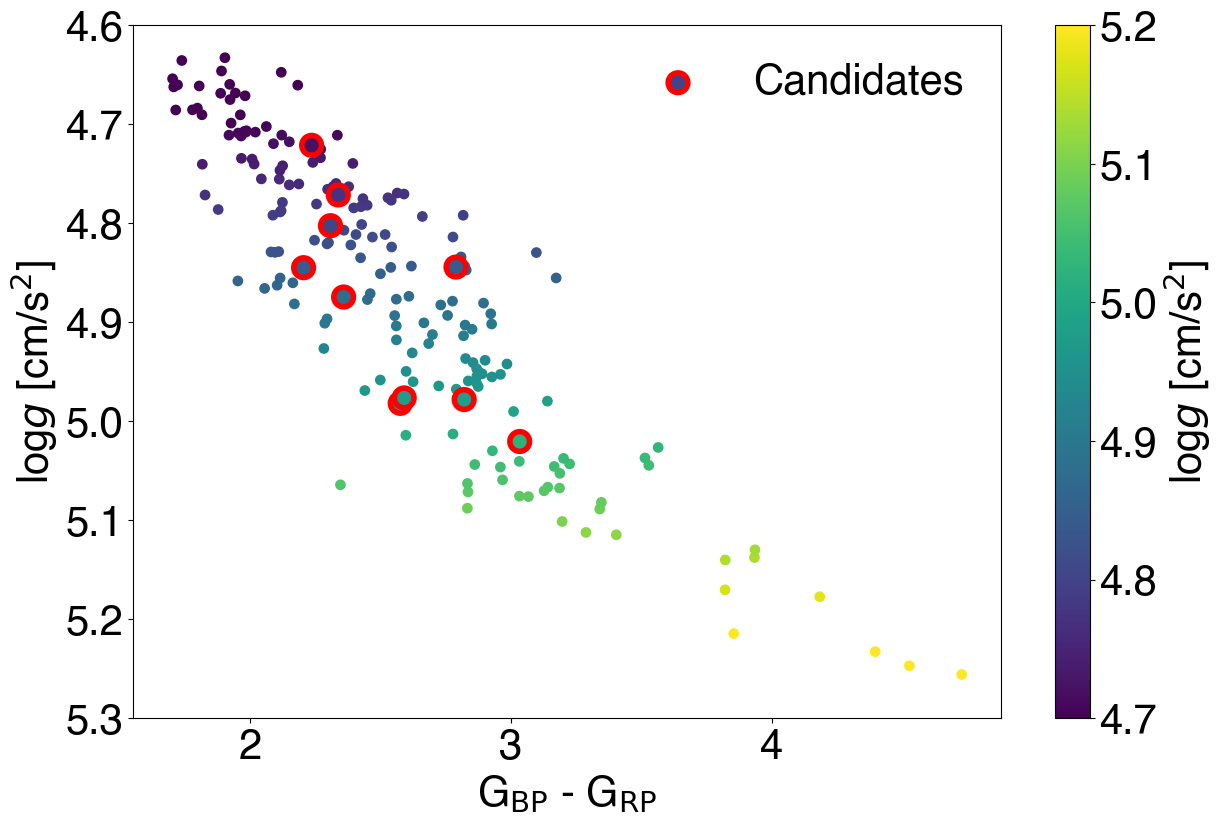}
    \caption{Surface gravity vs.\  $G_{\rm BP} - G_{\rm RP}$ colour-coded according to surface gravity. We use red circles to denote \citetalias{suazo24} candidates and include stars from the \citet{mann2015} sample (small bullets). All of our candidates follow the main sequence, i.e.\ surface gravity increases towards later spectral (sub)type.} 
    \label{fig:surface_color}
\end{figure}

Knowing stellar masses and radii enables the determination of surface gravity values for each star. Young stars, with the same mass distributed over a larger radius than for their older main-sequence counterparts, exhibit smaller surface gravity values, providing a diagnostic indicator of youth. In Figure~\ref{fig:surface_color}, the surface gravity plotted against colour demonstrates the agreement between our candidates and standard main-sequence M dwarfs, with background stars representing main-sequence M dwarfs from the \citet{mann2015} sample. No significant deviations from typical main-sequence M dwarfs are observed.

\begin{figure}
    \centering
    \includegraphics[width=\columnwidth]{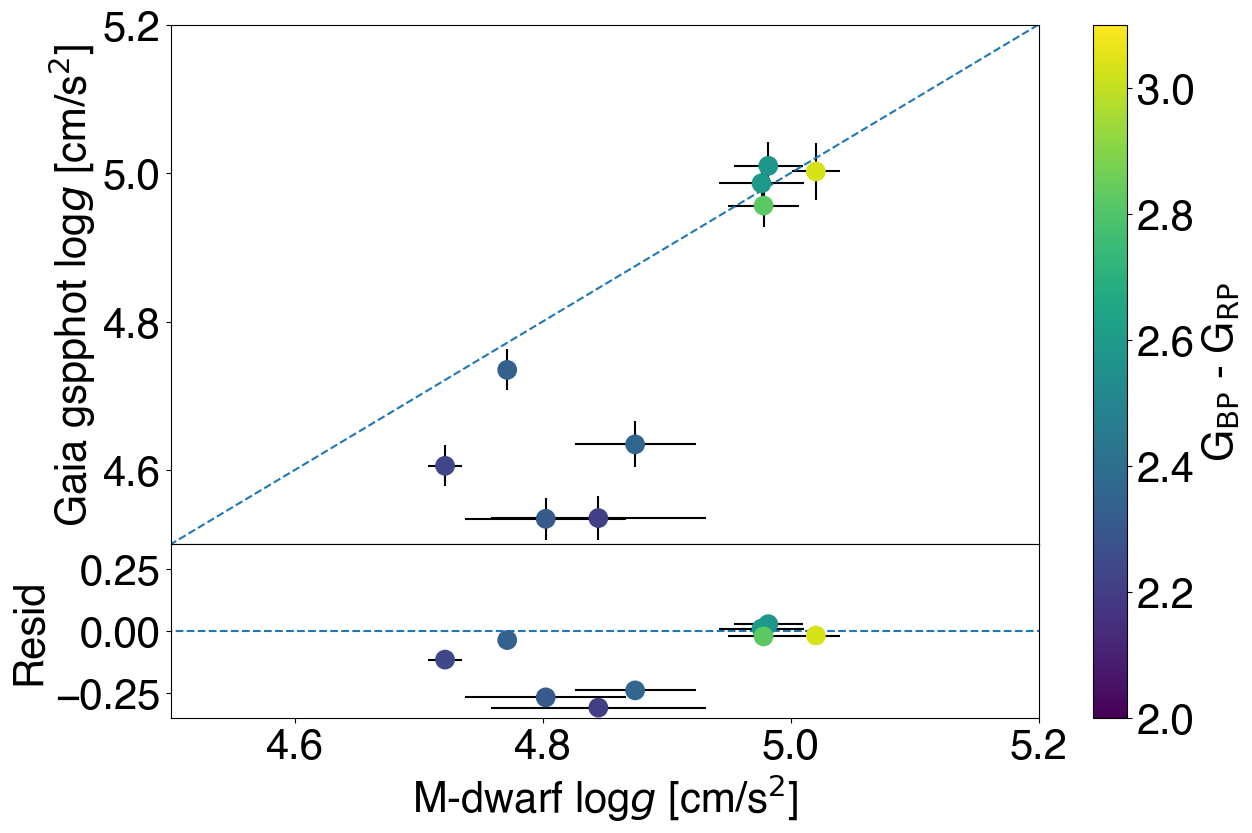}
    \caption{Surface gravity values derived from the Gaia DR3 GSP-phot pipeline vs.\ surface gravity by employing masses and stellar radii derived from empirical relations for our sources. The colour coding represents $G_{\rm BP} - G_{\rm RP}$. Both independent relations agree for the late M dwarfs, while Gaia underestimates the surface gravity for the earliest members of the sample.}
    \label{fig:surface}
\end{figure}

We also compare the surface gravities with the ones provided by the GSP-Phot module \citep{gaia_apsis} as part of Gaia DR3. This module derives stellar parameters from modeling the low-resolution BP/RP Gaia spectra. Figure~\ref{fig:surface} compares the values provided by Gaia and those calculated from the stellar masses and radii from the previous sections. We notice that the redder stars match the one-to-one relation very well, whereas, for the bluer stars, Gaia underestimates these values compared to the one computed using stellar mass and radius. However, the average offset is not large ($\simeq 0.15$\,dex).

\subsubsection{Metallicities}
\label{sec:metal}

Stellar metallicity refers to the relative abundance of elements beyond helium, with the iron abundance ratio [Fe/H] commonly employed as a proxy for overall metallicity. Assessing the metallicity of M dwarfs is a challenging task, primarily because of the complexities associated with their spectra. 
A multitude of lines arising from various molecular species (such as TiO, VO, ZrO, FeH, CaH in the optical regime; H2O, CO in the near-infrared) dominate the spectrum, complicating the determination of atmospheric parameters \citep[e.g.,][]{allard1997,delaverny2012,passegger2018,marfil2021}. 

Deriving stellar metallicity is a complex task, and we adopt two independent empirical relations for our analysis. Firstly, we utilize the empirical relation proposed by \citet{rains2021}, which follows the methodologies of \citet{johnson2009} and \citet{schlaufman2010}. In this method, metallicities are estimated by determining the color offset of a star relative to the mean main sequence in the $M_{\rm K_{s}}$-(colour) space. While \citet{johnson2009} and \citet{schlaufman2010} utilized the ($V - K_{s}$) colour to trace the mean main sequence, \citet{rains2021} opt for the $BP - K_{s}$ colour.

\begin{figure}
    \centering    \includegraphics[width=\columnwidth]{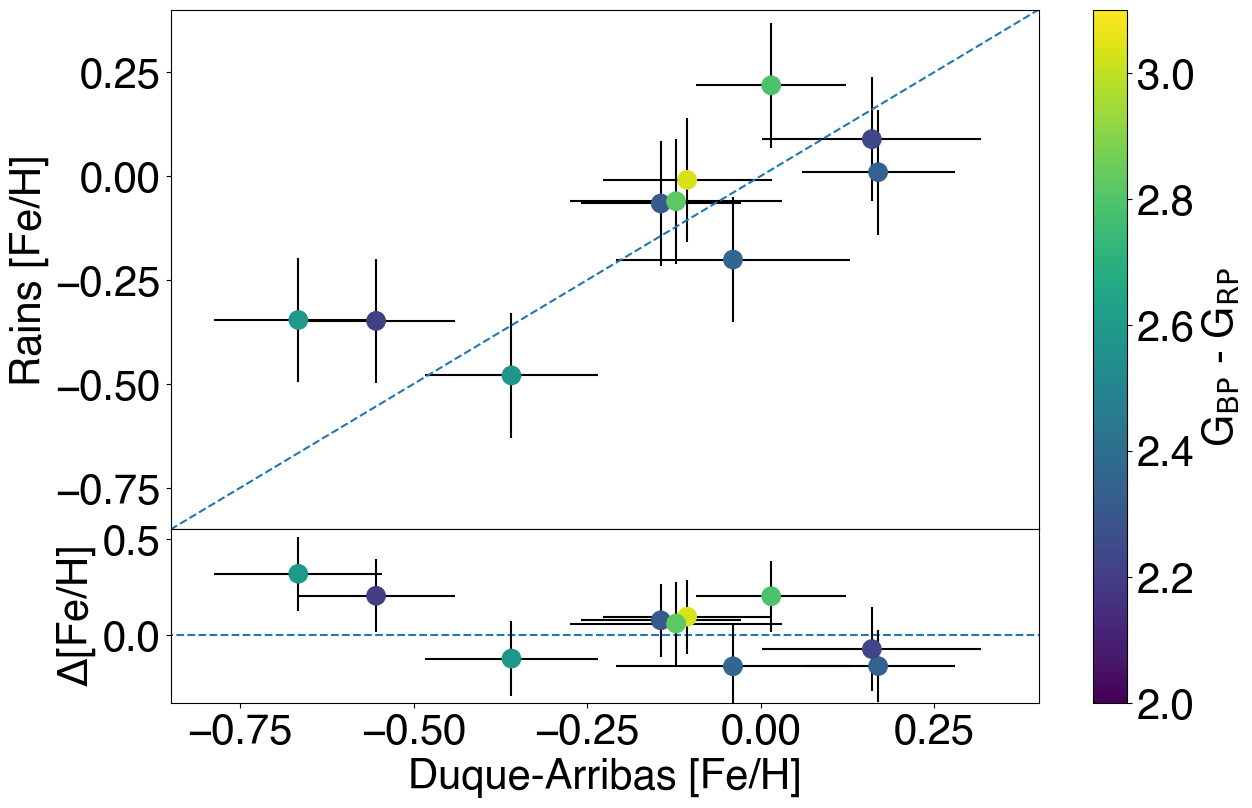}
    \caption{Metallicities derived using \citet{rains2021} empirical relations vs metallicities derived using fitting function from \citet{duque2023}. The colour bar represents $G_{\rm BP} - G_{\rm RP}$. The two independent relations are in good agreement.}
    \label{fig:metal}
\end{figure}

Concurrently, we employ another set of empirical relations derived by \citet{duque2023}. Their study focused on identifying the combination of colors and absolute magnitudes that yield the most accurate metallicity predictions. They incorporated not only optical colours but also WISE infrared colours in their analysis. Their analysis led to the finding that the most suitable fitting function is a polynomial function that utilizes the absolute magnitude in the $G$ band, the Gaia $G_{\rm BP} - G_{\rm RP}$ colour and the CATWISE \citep{catwise} $W1 - W2$ colour as input parameters. Figure~\ref{fig:metal} showcases the metallicity values obtained by using the above-mentioned empirical relations. We notice that the metallicity estimates are in agreement and scatter around solar metallicity, with the exception of two stars that are too metal-poor for the \citet{duque2023} empirical relations to hold. 

\subsubsection{Effective stellar temperatures}

Accurate effective-temperature determinations are crucial as $T\rm _{eff}$ provides insight into the spectral energy distribution and luminosity of a star and thus influences the location of the habitable zone in a stellar system \citep{Kasting1993}.

\begin{figure}
    \centering
    \includegraphics[width=\columnwidth]{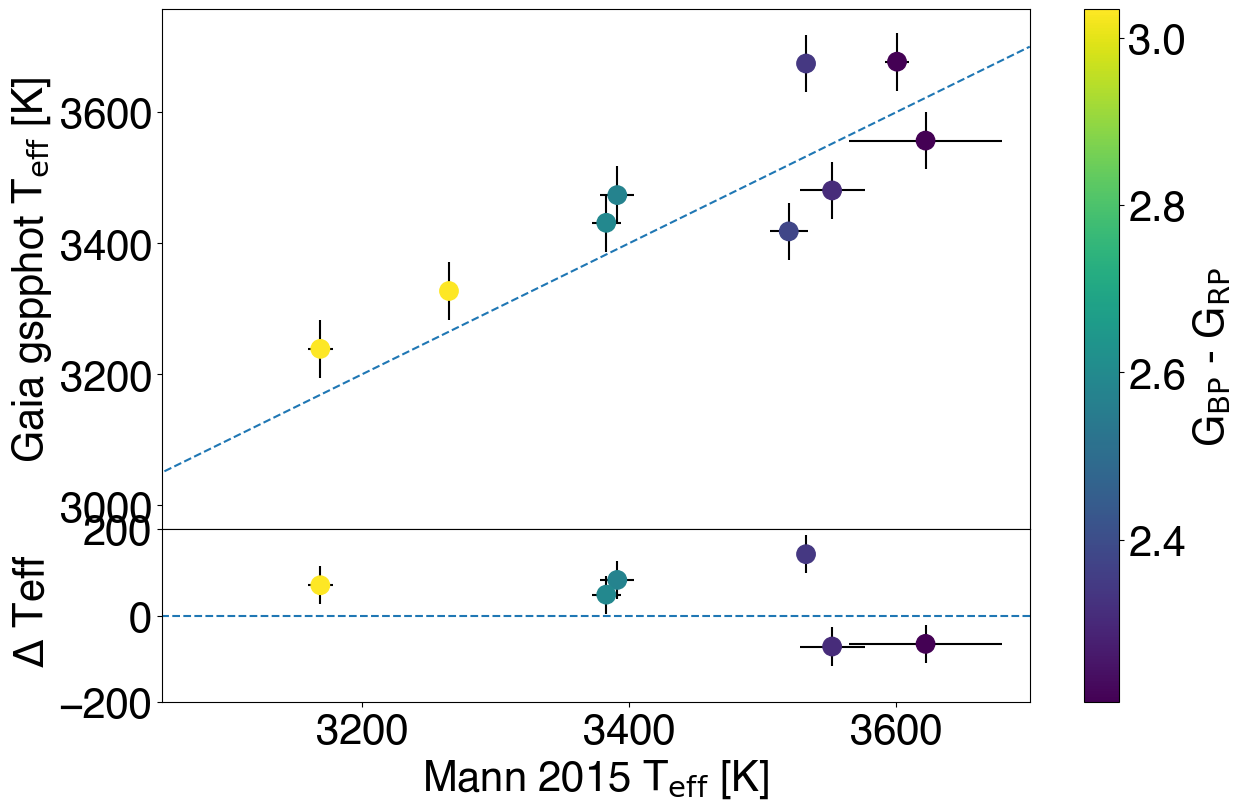}
    \caption{Effective temperatures from GSP-Phot compared to those derived from the \citet{mann2015} empirical relation. Colour represents Gaia $\rm G_{BP} - G_{RP}$. GSP-Phot slightly overestimates the effective temperatures on average, but the offset lies within the range expected from Gaia \citep{Andrae2023}.}
    \label{fig:temperatures}
\end{figure}

To determine stellar temperatures, we applied one of the photometric relations of \citet{mann2015}, specifically effective temperatures as a function of Gaia $G_{\rm BP} - G_{\rm RP}$ colour and metallicity. However, we used a re-derived version of the \citet{mann2015} relation compatible with Gaia DR3 data (Rains, private communication) and employed \citet{rains2021} metallicities to estimate effective temperatures. Although both relations employed in Section~\ref{sec:metal} produced similar results, the \citet{rains2021} relation covers a broader colour and metallicity range. The effective temperatures derived from this procedure are presented in Table~\ref{tab:stellar}. Furthermore, Figure~\ref{fig:temperatures} compares these effective temperatures values with those obtained from the GSP-Phot in Gaia DR3. We notice a slight offset of $\leq 100$ K of Gaia-based effective temperatures compared to those derived using the \citet{mann2015} empirical relation. 

\begin{table*}
\caption{Stellar parameters. Sources for metallicity estimates: $^{\rm a}$ \citet{duque2023}, $\rm ^{b}$ \citet{rains2021}.}
\centering
\begin{tabular}{c|c|c|c|c|c|c|c|c|c|c|c|c}
\hline
Label & Gaia DR3 ID & $M$ [M$_{\odot}$] & $R$ [R$_{\odot}$] & $T\rm _{eff}$ [K] & log$g$ [cm/s$^2$] & [Fe/H]$^{\rm a}$ & [Fe/H]$^{\rm b}$ \\
\hline
A & 3496509309189181440 & 0.374 $\pm$ 0.009 & 0.383 $\pm$ 0.011 & 3318 $\pm$ 44 & 4.84 $\pm$ 0.02 & ~\,~0.015 $\pm$ 0.107 & ~\,~0.218 $\pm$ 0.150 \\
B & 4843191593270342656 & 0.225 $\pm$ 0.009 & 0.255 $\pm$ 0.007 & 3321 $\pm$ 44 & 4.97 $\pm$ 0.03 & $-$0.666 $\pm$ 0.121 & $-$0.346 $\pm$ 0.150 \\
C & 4649396037451459584 & 0.190 $\pm$ 0.012 & 0.223 $\pm$ 0.006 & 3166 $\pm$ 44 & 5.02 $\pm$ 0.03 & $-$0.106 $\pm$ 0.121 & $-$0.009 $\pm$ 0.150 \\
D & 2660349163149053952 & 0.220 $\pm$ 0.009 & 0.250 $\pm$ 0.007 & 3305 $\pm$ 44 & 4.98 $\pm$ 0.03 & $-$0.359 $\pm$ 0.124 & $-$0.479 $\pm$ 0.150 \\
E & 3190232820489766656 & 0.373 $\pm$ 0.013 & 0.382 $\pm$ 0.011 & 3559 $\pm$ 44 & 4.84 $\pm$ 0.02 & $-$0.554 $\pm$ 0.114 & $-$0.348 $\pm$ 0.150 \\
F & 2956570141274256512 & 0.466 $\pm$ 0.012 & 0.465 $\pm$ 0.013 & 3533 $\pm$ 44 & 4.77 $\pm$ 0.02 & ~\,~0.169 $\pm$ 0.110 & ~\,~0.009 $\pm$ 0.150 \\
G & 2644370304260053376 & 0.428 $\pm$ 0.012 & 0.430 $\pm$ 0.012 & 3540 $\pm$ 44 & 4.80 $\pm$ 0.02 & $-$0.144 $\pm$ 0.114 & $-$0.066 $\pm$ 0.150 \\

H & 2437221214075471744 & 0.336 $\pm$ 0.013 & 0.350 $\pm$ 0.010 & 3483 $\pm$ 44 & 4.87 $\pm$ 0.03 & $-$0.040 $\pm$ 0.169 & $-$0.200 $\pm$ 0.150 \\
I & 3854090071297359616 & 0.223 $\pm$ 0.007 & 0.253 $\pm$ 0.007 & 3254 $\pm$ 44 & 4.97 $\pm$ 0.02 & $-$0.112 $\pm$ 0.152 & $-$0.060 $\pm$ 0.150 \\
J & 651765552072217216 & 0.524 $\pm$ 0.013 & 0.522 $\pm$ 0.015 & 3616 $\pm$ 44 & 4.72 $\pm$ 0.02 & ~\,~0.160 $\pm$ 0.157 & ~\,~0.088 $\pm$ 0.150 \\
\hline
\end{tabular}
\label{tab:stellar}
\end{table*}

\begin{figure}
    \centering    \includegraphics[width=\columnwidth]{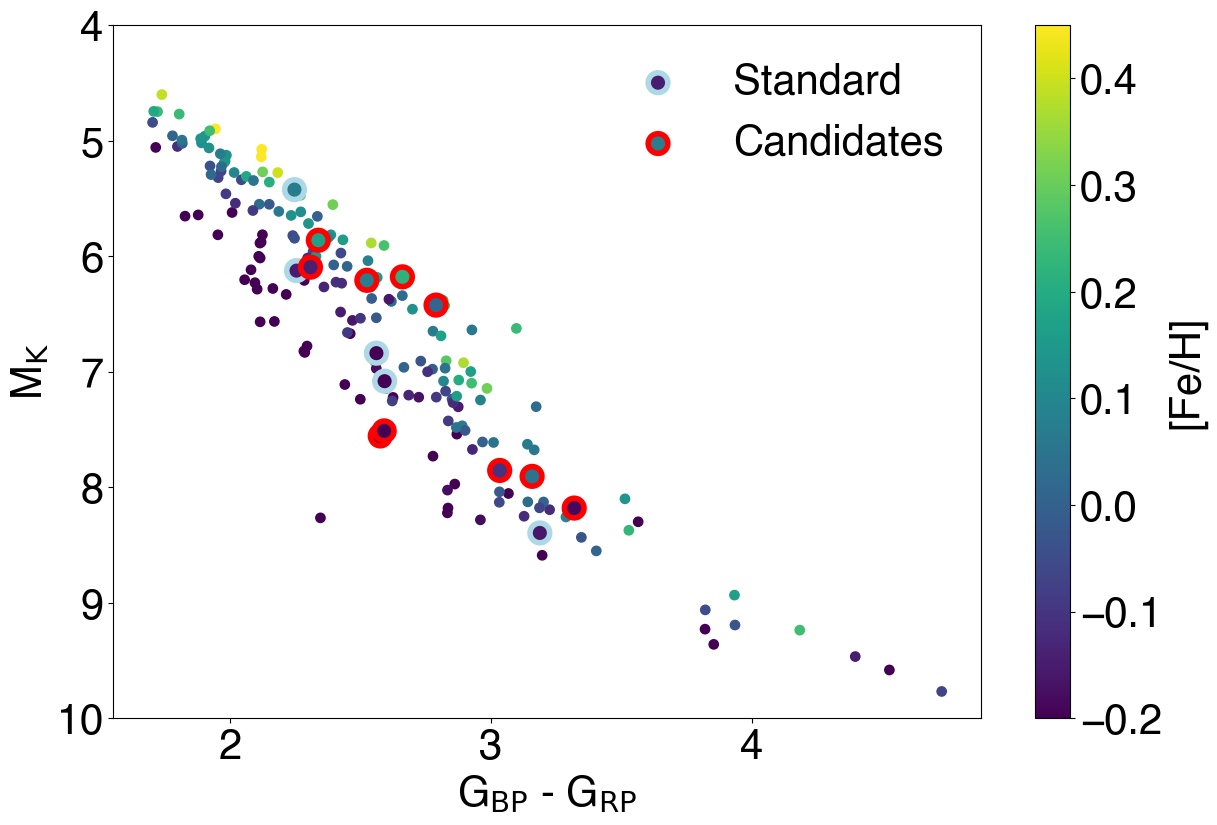}
    \caption{Colour-magnitude diagram showing our candidates (red circle), standard stars with ALFOSC spectra (light-blue circle) and M dwarfs from the \citet{mann2015} sample (smaller bullets). Our stars follow the main sequence showcasing also the relation between colour and metallicity.}
    \label{fig:cmd_feh}
\end{figure}

\subsection{Spectroscopy}
\label{sec:spec}

\begin{figure}
    \centering
    \includegraphics[width=\columnwidth]{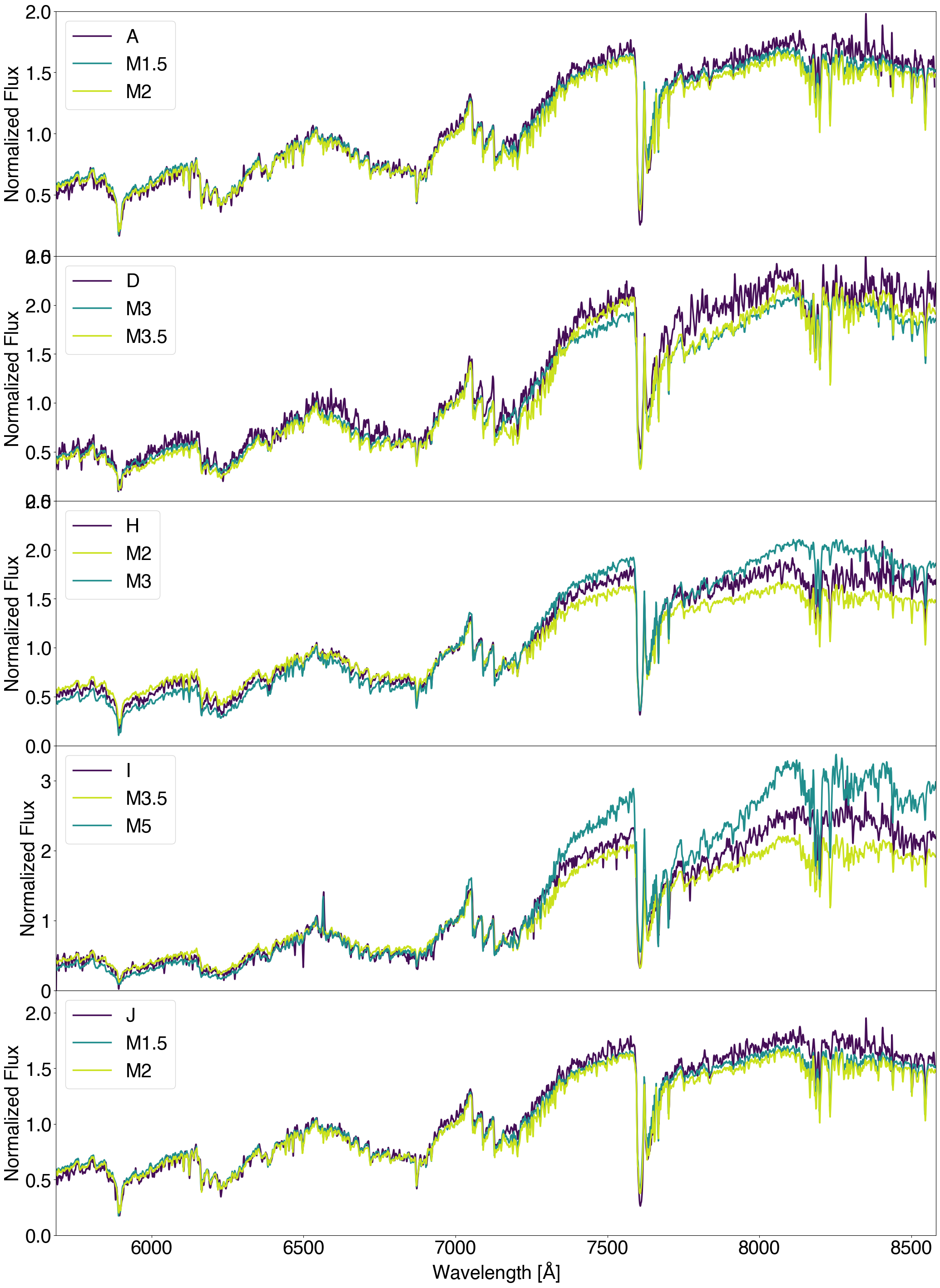}
    \caption{Stellar spectra for candidates A, D, H, I, and J along with standard main-sequence M dwarfs to facilitate comparison. There is a fair match between our standard stars and the optical spectra of our candidates. No H$\alpha$ in emission is observed in all but one case.}
    \label{fig:spectra}
\end{figure}

In addition to the photometric analysis, we conducted a spectroscopic analysis to gather more information about our candidates. As mentioned in Section~\ref{sec:observations}, we obtained low-resolution spectra ($\lambda/\delta\lambda \sim$ 2000) in the $R$ band using the ALFOSC spectrograph at the NOT. Spectra were collected for candidates A, D, H, I and J (and the standard stars of Table \ref{tab:standard}). Figure~\ref{fig:spectra} displays the calibrated spectra for these five sources and some of the standard stars to facilitate the comparison with main-sequence M dwarfs.

One prominent feature observed in the spectra of candidates A, D, H, and J is the absence of H${\alpha}$, confirming that these sources are not pre-main-sequence stars. Furthermore, the spectra of these four candidates closely match those of our standard stars. In the case of candidate I, H${\alpha}$ is observed in emission. However, this feature is also present in the standard star GJ 83.1. Utilizing the H${\alpha}$ Equivalent Width (EW) definition by \citet{west2011}, we found that the H${\alpha}$ EW for candidate I is approximately $-2.8$ Å, while H${\alpha}$ EW for the standard star GJ 83.1 is around $-2.3$ Å (the negative sign indicating emission). Various empirical relations \citep[e.g.,][]{barrado2003,white2003} suggest that this emission is consistent with stellar activity rather than gas emission from circumstellar material. More specifically, a young star is considered accreting if H$\alpha$ EW $<$ $-$20 Å for M3 - M5.5 \citep{white2003}. In our reference star and candidate I, the values are well above this threshold. Both empirical relations also indicate that H$\alpha$ becomes more common in late-type M dwarfs, which is in line with detecting it in the coolest star in our sample of stars with available spectra. 

\subsection{Infrared disk taxonomy}
\label{sec:infrared}

As previously mentioned, circumstellar material could explain the spectral energy distributions of our sources. Considering the possibility of circumstellar disks implies considering a full family of primordial disks. Here, we employ \citet{espaillat2012} taxonomy of each evolutionary stage, which is summarized by \citet{murphy2018} as follows: Full disks are optically thick at near- and mid-infrared wavelengths and primordial dust and gas have not been cleared yet. Pre-transitional \citep{espaillat2007} and transitional disks have large inner gaps or holes, respectively. Evolved disks are in the stage when the circumstellar material starts becoming optically thin, but does not possess large holes or gaps. Evolved transitional disks are optically thin and have large holes. All of the above classifications can be considered primordial disks, whereas debris disks are composed of second-generation dust generated by collisions of planetesimals after the primordial disk has dissipated.

\begin{figure}
    \centering
    \includegraphics[width=\columnwidth]{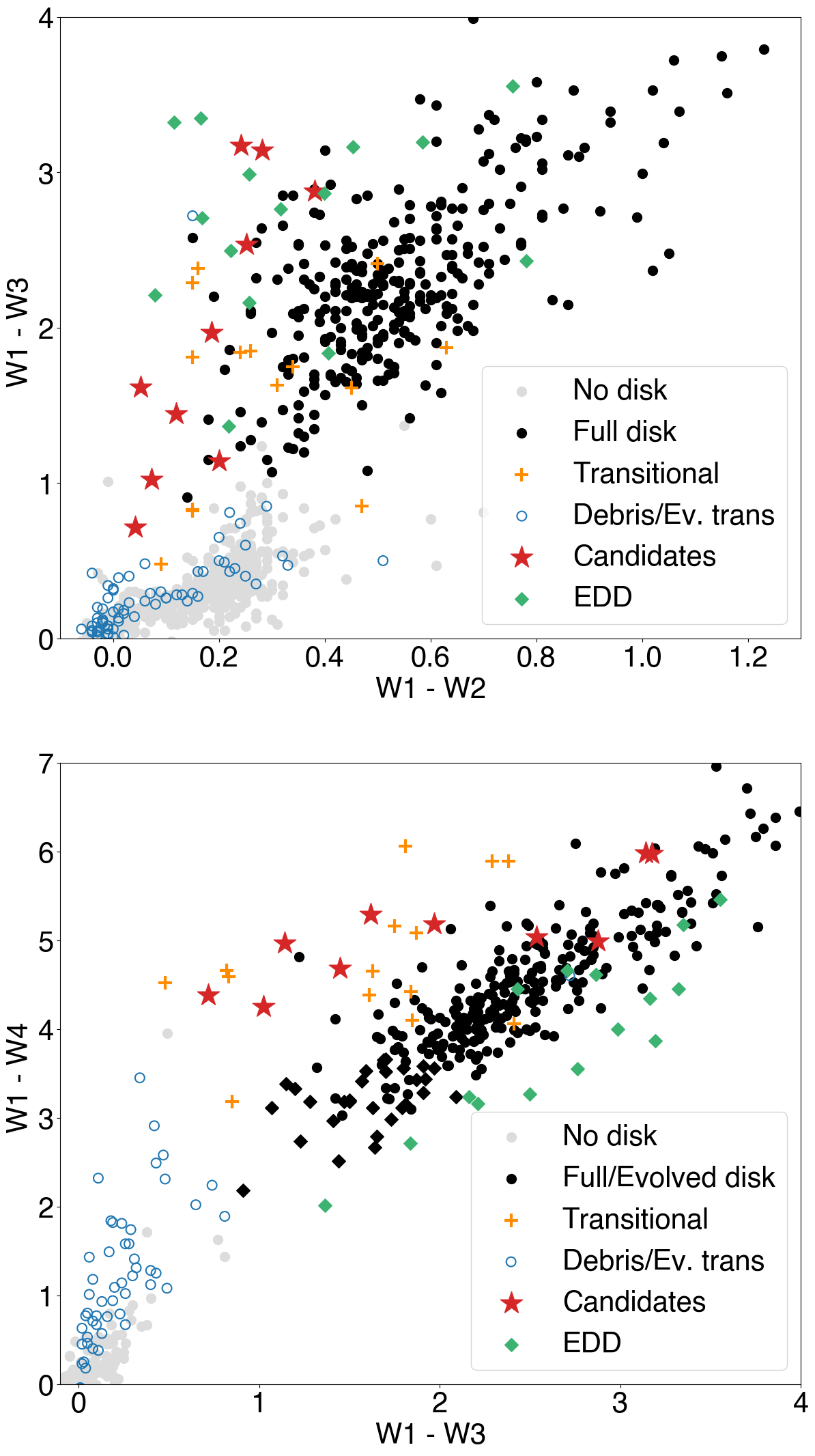}
    \caption{AllWISE colour-colour diagrams comparing our candidates to stars in different evolutionary stages in the Taurus and Upper Sco star-forming regions from the studies of \citet{Esplin2014} and \citet{Luhman2012}, respectively. We also add Extreme Debris Disks (EDD) from \citet{moor21}.
    Each colour/symbol represents a different evolutionary stage. Colour-wise, our sources are not compatible with full or debris disks, but they lie in an intermediate regime partly overlapping with transitional disks.}
    \label{fig:infrared}
\end{figure}

To compare our candidates with all stages of primordial-disk evolution, we conducted a similar analysis to the one developed by \citet{murphy2018} investigating the location of our candidates in AllWISE colour-colour diagrams. In Figure~\ref{fig:infrared}, we plot two AllWISE colour-colour diagrams showcasing the distribution of different evolutionary stages for stars in the Upper Scorpius and Taurus star-forming regions from \citet{Luhman2012} and \citet{Esplin2014}. As can be seen in both colour-colour diagrams, different evolutionary stages are located across various regions, from full disks in the upper right corner to no disks in the lower left corner. Additionally, we included 15 stars with EDDs \citep{balog09} from the \citet{moor21} sample, which comprises decades of research in this relatively new field. EDDs are characterized by their high fractional luminosities ($f > 0.01$) and by dust-rich warm debris ($T_{\rm dust} > 300 $K). It is noteworthy that all EDDs discovered up to date have been found around FGK-type stars.

By comparing the AllWISE colours of our candidates, we see no clear preference for any particular category. According to \citetalias{suazo24} models, these candidates do not possess any excess in $W1$ nor in $W2$, but they do simultaneously in $W3$ and $W4$ (more prominently in $W4$), hence explaining the location of our sources in this diagram. In the upper panel, our candidates remain on the left-hand side of the diagram since there are no $W1$ or $W2$ excesses, while the evident $W4$ excess explains most of the crowding in the upper region in the lower panel. We notice that our sources resemble transitional disks colour-wise, additionally, the fractional luminosity is compatible with this evolutionary stage.

\section{Discussion}
\label{sec:discussion}

This work characterizes and analyzes ten M dwarfs exhibiting unusual infrared radiation. We analyze the seven sources from \citetalias{suazo24} and add three additional sources since we have followed them up with ALFOSC. We derive stellar parameters from empirical relations and compare them with the ones provided by Gaia DR3 \citep{gaiadr3_apsis1}.

When deriving the stellar parameters of our sample, we find some offsets between the empirical relations used in this work and the stellar parameters provided by the Apsis module GSP-Phot  \citep{gaiadr3_apsis1}, in particular for $T_{\rm eff}$. This offset was previously noted by the Gaia team \citep{gaiadr3_apsis1,Andrae2023}. Stellar parameters such as log$g$ and [M/H] affect the $G_{\rm BP}$ and $G_{\rm RP}$ spectra only weakly compared to $T_{\rm eff}$, so when comparing Gaia results with those reported by other surveys, e.g., LAMOST \citep{wu2011,wu2014}, GALAH \citep{buder2021}, Gaia-ESO \citep{gilmore2012,Gilmore2022} and RAVE \citep{steinmetz2020}, GSP-Phot stellar parameters are often found to be offset from the values inferred by the above-mentioned surveys. Temperature-wise, GSP-Phot results result in a median absolute error in $T\rm _{eff}$ of 119 K and a mean absolute error of 180 K across the various data sets analysed\citep{Andrae2023}. 

\citet{Andrae2023} also compares the surface-gravity values from the GSP-Phot module with that of the above-mentioned surveys and notices an agreement with those values within 0.25 dex. Additionally, \citet{Andrae2023} compares these surface gravity values with the ones obtained from asteroseismic studies \citep{Serenelli2017,yu2018}, and this comparison reveals that log$g$ uncertainties from GSP-Phot are underestimated up to a factor 10.
 
Utilizing recent and independent empirical relations, we estimated iron abundances using \citet{duque2023} and \citet{rains2021}. Both relations agree for our candidates to within 1.2$\sigma$ (derived from the mutual error bars). Discrepancies get larger outside the validity  range for these relations. The \citet{rains2021} relation is derived from a sample of 69 stars with metallicities within the range $-$1.0 $<$ [Fe/H] $<$ 0.5, and it applies to stars where Gaia colours range from 1.51 $< G_{\rm BP} - G_{\rm RP} <$ 3.3. On the other hand, \citet{duque2023} uses a narrower range of metallicities ($-$0.5 $<$ [Fe/H] $<$ 0.5), but their sample is much larger, comprising ~5000 stars.

Although Gaia DR3 provides metallicities \citep{gaiadr3_apsis1}, it is known that GSP-Phot estimates are typically biased by $-$0.2\,dex and, similar to surface gravities, this parameter is affected by additional systematics. \citet{gaiadr3_apsis1} suggest caution when using GSP-Phot metallicity estimates without further investigation. 
Notice that Gaia DR3 provides bulk metallicities [M/H] compared to the iron abundances [Fe/H] estimated in  ection~\ref{sec:metal}. 
We acknowledge the power of using empirical relations, especially the ones used in this work that make use of infrared photometry. Information in this region of the electromagnetic spectrum becomes very valuable when estimating metallicities compared to optical photometry only.

Temperature-wise, the GSP-phot module either overestimates or underestimates the effective temperature compared to the empirical relation of \citet{mann2015}. However, the deviations match the mean absolute error (180 K) found by \citet{Andrae2023} when comparing effective-temperature estimates from GSP-Phot to those of different surveys. We thus find no significant offsets in $T\rm _{eff}$ values between the  empirical relations and those provided by Gaia DR3.

\citet{Andrae2023} typically compare Gaia results to literature results for FGK-type stars. Although M dwarfs are usually not considered in the comparison, the astrophysical parameters derived from empirical relations here agree with the deviations found for early-type stars. However, in many cases, Gaia-parameter  uncertainties tend to be smaller than literature or empirical-relations values, which, as claimed by \citet{gaiadr3_apsis1}, suggests that random errors do not drive these deviations but they are rather dominated by systematic errors (such as different temperature scales). Nevertheless, regardless of the differences between the astrophysical parameters derived in Section~\ref{sec:photometry} and the ones provided by Gaia DR3, they all agree on the fact that our stars are indeed main-sequence M dwarfs making it very unlikely that these sources to be pre-main sequence stars.

Our spectroscopic analysis reveals no H$\alpha$ emission in four of our five stars with available spectra. This ultimately rules out the possibility of a hydrogen-dominated accretion disk being heated up, thereby emitting H${\alpha}$ photons \citep{barrado2003} and infrared radiation. However, for candidate I and the standard star GJ 83.1, we see H$\alpha$ in emission. For late-type stars such as M dwarfs, H${\alpha}$ emission can be produced from stellar activity rather than an accretion disk. Various works \citep[e.g.,][]{barrado2003,white2003} have been able to distinguish H$\alpha$ originating from either stellar activity or an accretion disk. The H$\alpha$ equivalent width can be used to separate accretors from active stars. \citet{barrado2003} and \citet{white2003} developed empirical relations for this separation. Both works correlate the threshold for this separation with the spectral type and subclass of the stars. For the spectral type of our H$\alpha$ emitters, both stars lie well above that threshold (smaller H$\alpha$ EW) to categorize a star as an accreting source \citep[< $-$10 Å;][]{white2003}, where the H$\alpha$ equivalent widths of candidate I and GJ 83.1 are above $-$3 Å.

Stellar ages are crucial parameters providing insight into various astrophysical phenomena, including planetary and Galactic evolution. In the case of stars potentially hosting warm debris material, stellar ages can help determine the origin of such debris disks. However, without extra information (e.g. asteroseismology) stellar ages can only be estimated, not precisely determined, especially among main-sequence stars that are characterized by their slow H-burning evolution.

As all our sources are TESS sources, we could obtain light curves to be used for age estimation through gyrochronology, a technique employing the rotational period as a tracer of stellar age \citep{Barnes2003}. As stars age, their rotational period slows down due to magnetic braking \citep{skumanich1972} which leads to the loss of angular momentum \citep{Barnes2007}. Although various calibrations of stellar rotation to stellar ages exist in the literature, we encountered challenges in determining ages.

We used the \texttt{eleanor} \citep{eleanor} Python package to obtain TESS light curves from the Full Frame Images (FFI). This algorithm automatically locates the target in the FFI, builds a time series of postcards containing the pixels with the target and immediate background, creates a target pixel file (TPF), identifies centroid shifts in the time series, builds the light curve applying a chosen aperture, and performs correction for systematics. We obtained light curves for all the sectors on which our targets were observed, where a sector corresponds to a region of the sky observed during $\sim $27 days. We then applied the Lomb-Scargle periodogram \citep{Lomb1976,Scargle1982} from the Astropy package to study the periodicities of the light curves and determine if they are rotationally modulated. We found no significant periodicities in the light curves for any of our targets. Candidates A, E, and F have TESS light curves from two sectors separated by two years, which makes it impossible to compute a reliable rotation period even if the photometry is rotationally modulated. It is important to notice that one TESS sector being about 27 days long, it is challenging to detect rotation periods longer than $\sim$14 days, particularly if no consecutive sectors are observed.

Finally, we compared the AllWISE colours of our candidates with those of different early evolutionary stages (Fig~\ref{fig:infrared}). We see our candidates possibly resembling transitional disks: disks in the process of clearing their inner regions \citep{strom1989}. Because of the clearing phenomenon, disks with cavities of different sizes and located at different distances from the central star have been discovered \citep[e.g.,][]{calvet2002,calvet2005,dalessio2005,brown2007}. SED-wise, transitional disks are very complex due to the intricate nature of possible gap distributions. However, they can still be characterized by some significant features. SEDs of transition disks usually show no excess emission in the near and mid-infrared and an excess at wavelengths $>$ 20 $\mu$m \citep{calvet2005}, which indicates no hot/warm dust close to the star. Since they are evolving from full disks, they still bear high dust content and have high fractional luminosities \citep[e.g.][]{Michel2021} which makes this type of source compatible with our candidates. Just like in this study, in various others \citep[e.g.;][]{esplin2018,murphy2018}, optical spectroscopy has helped to contribute to our understanding of the nature of infrared sources. \citet{esplin2018} surveyed the Upper Sco association using mid-infrared photometry from WISE aimed to detect all types of disks described in Section~\ref{sec:infrared} including transition disks. Aided with optical spectroscopy when possible, \citet{esplin2018} confirm various circumstellar candidates from lithium absorption features that are tracer of youth \citep[e.g.;][]{martin1997,martin1998}, as well as weak sodium absorption lines that are very sensitive to surface gravity \citep{Schlieder2012}. Additionally, optical spectroscopy in transition disks has shown that accretion is ongoing, and therefore, H$\alpha$ is emitted as a consequence of this process \citep{sicilia2008,esplin2018}. Stellar variability is also a characteristic feature of circumstellar disks, and it has been used to determine the nature of transition disks \citep[e.g.][]{nagel2021}. Although features such as the colour and the fractional luminosity of our sources are compatible with transition disks, the lack of other youth indicators, such as variability \citepalias{suazo24} and H$\alpha$ in emission, seems to indicate that our sources are either not transition disks or somewhat anomalous members of this category. 

The blackbody nature of the infrared excess observed in our Dyson-sphere candidates can be further studied with existing telescopes that either probe the spectral region close to the peak, or in the Rayleigh-Jeans tail, of this blackbody spectral energy distribution. Such measurements would both allow an independent verification of the infrared excess, better constrain the blackbody temperature and check for deviations from a pure blackbody spectrum (for instance due to mid-infrared emission features due to cosmic dust). In scenarios where the infrared excess originates from a background object (most likely a heavily dust-obscured galaxy or active galactic nucleus) projected onto the star seen at optical and near-infrared wavelengths, either spectroscopy or high-resolution imaging at longer wavelengths may confirm the presence of this second source (by inferring that the sources is extended as opposed to point-like in the infrared, or by detecting spectral features with a redshift that places it beyond the Milky Way). In fact, JWST Mid-Infrared Instrument (MIRI) observations of two of our candidates (D and E) have recently been analysed by \citet{Zackrisson2026} and reveal the presence of red background galaxies. Even radio observations can help to detect background galaxies, as \citet{Ren2026} show for candidate B.

\begin{figure*}
    \centering
    \includegraphics[width=0.7\textwidth]{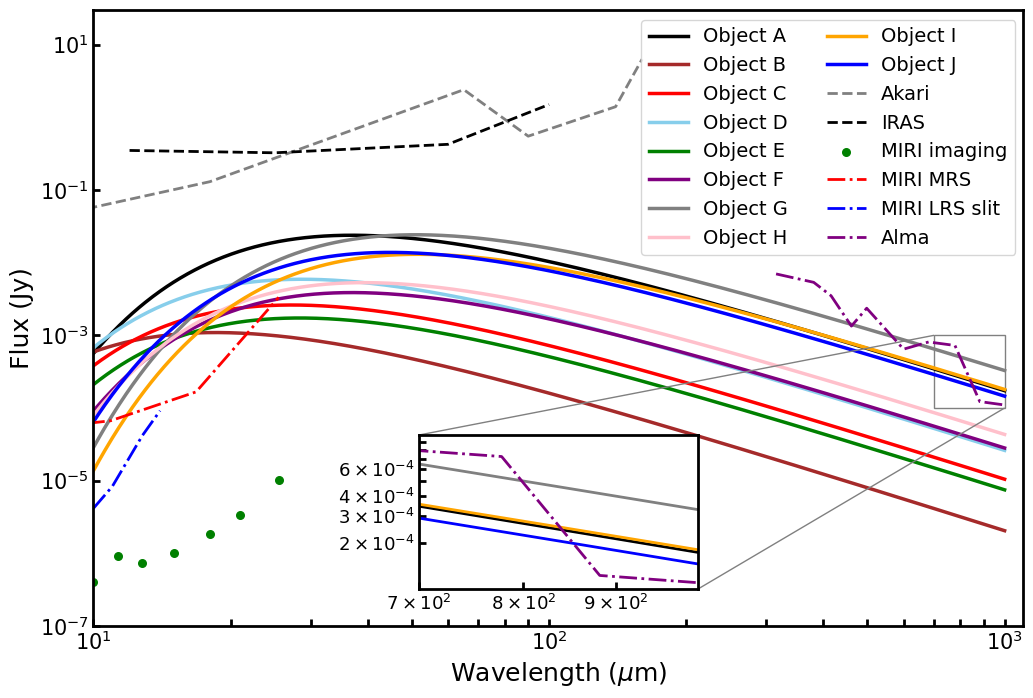}
    \caption{Blackbody spectra derived from the $T_{\rm eff}$ of the Dyson sphere candidates A to J throughout the wavelength range of 10 to 1000 microns together with the detection thresholds of telescopes Akari \citep{Murakami2007}, IRAS \citep{IRAS}, the ALMA 12$\sim$m array\protect\footnotemark and JWST/MIRI. In the case of JWST, both broadband imaging and spectroscopic limits (based on the LRS Slit or MRS\protect\footnotemark observations) are shown. In the case of JWST and ALMA, all limits refer to 5$\sigma$ confidence in 1 hour exposure time (where exposure times refer to observations per filter in the case of imaging; per band in the case of ALMA spectroscopy; per channel in the case of JWST/MIRI MRS spectroscopy and for the entire  plotted wavelength range in the case of JWST/LRS spectroscopy). As can be seen, all candidates are detectable throughout some part of the wavelength coverage of JWST. The inset shows a zoom-in on the part of the spectral energy distribution that may be detectable with ALMA for four of the candidates (A, G, I and J).}
    \label{fig:followup}
\end{figure*}

In Figure~\ref{fig:followup}, we explore the detectability of our Dyson sphere candidates in the 10--1000 $\mu$m range by showing the best-fitting blackbody SEDs of our Dyson sphere candidates, along with detection limits for Akari, Infrared Astronomical Satellite (IRAS), the JWST and the Atacama Large Millimeter/submillimeter Array(ALMA). All the Dyson-sphere candidates lie far below the photometric detection limits of the Akari and the IRAS all-sky surveys, but can readily be detected in imaging with JWST MIRI at $\approx 10$--26 $\mu$m. All objects are also expected to be sufficiently bright for spectroscopy using either JWST/MIRI Low-Resolution Spectroscopy (LRS) at wavelengths up to 14 $\mu$m. JWST/MIRI Mid-Resolution Spectroscopy (MRS) at $\approx 10$--26 $\mu$m range also remains a viable option, although all candidates may not be detectable throughout the entire wavelength interval. The Rayleigh-Jeans tail of the best-fitting blackbody spectral energy distributions mostly lie below the detection limits of ALMA, but four objects (candidates A, G, I and J) should be detectable at $\approx 900$--1000 $\mu$m.

\addtocounter{footnote}{-1}
\footnotetext{\url{https://almascience.eso.org/proposing/sensitivity-calculator}}
\addtocounter{footnote}{1}
\footnotetext{\url{https://jwst-docs.stsci.edu/jwst-mid-infrared-instrument/miri-performance/miri-sensitivity##MIRISensitivity-MRStimedependentsensitivity}}

\section{Conclusions}
\label{sec:conclusions}


An in-depth analysis of the seven Dyson-sphere candidates presented in \citetalias{suazo24}, along with three additional candidates, confirms that these sources are main-sequence M-dwarf stars. This situation leaves the enigma of the infrared excess unresolved for all these sources (but see below). All empirical relations used to re-derive the stellar parameters are consistent with the hypothesis of main-sequence stars. GSP-Phot results from Gaia DR3 also confirm this picture.

We spectroscopically confirm this hypothesis for five sources in which the absence of high levels of H$\alpha$ in emission indicates that our sources are not young, as also suggested by the high surface gravities derived from empirical relations. When comparing our sources with the entire taxonomy of circumstellar disks, we find that they broadly resemble transition disks. However, in the absence of tell-tale youth indicators this seems an unlikely explanation, at least for the five stars with follow-up spectroscopy. 

Since this work was finished, several of our candidates have been followed up: see the careful centroid analysis and archival-data search of \citet{Ren2026} for candidates B and C revealing background objects in radio and IR, and the JWST/MIRI observations of \citet{Zackrisson2026} for candidates D and E. These follow-up studies now make red background galaxies the leading hypothesis for the IR excess of the seven (plus three) dwarfs selected among 5 million nearby stars cross-matched between Gaia and WISE. These ``needle in the haystack'' objects are star-galaxy superpositions unresolvable at the limited spatial resolution of WISE. Additional observations capable of confirming this scenario are already underway. 

\section{Acknowledgement}
Based on observations made with the Nordic Optical Telescope, owned in collaboration by the University of Turku and Aarhus University, and operated jointly by Aarhus University, the University of Turku and the University of Oslo, representing Denmark, Finland and Norway, the University of Iceland and Stockholm University at the Observatorio del Roque de los Muchachos, La Palma, Spain, of the Instituto de Astrofisica de Canarias. The NOT data were obtained under program ID 67-704.

This work has made use of data from the European Space
Agency (ESA) mission Gaia 
(\url{https://www.cosmos.esa.int/gaia}), processed by the Gaia Data Processing and Analysis Consortium (DPAC, 
\url{https://www.cosmos.esa.int/web/gaia/dpac/consortium}). Funding for the DPAC has been provided by national
institutions, in particular the institutions participating in the Gaia Multilateral Agreement.

This publication makes use of data products from the Wide-field
Infrared Survey Explorer, which is a joint project of the University of California, Los Angeles, and the Jet Propulsion Laboratory/California Institute of Technology, funded by the National Aeronautics and Space Administration.

This work was supported (in part) by the European Union (ChETEC-INFRA, project no. 101008324).

PCZ acknowledges support from the STFC consolidated grant number ST/V000861/1, UKSA grant number ST/X002217/1, UKRI/ERC Synergy Grant EP/Z000181/1 (REVEAL), and the L’Oréal-UNESCO For Women in Science UK and Ireland Young Talent Awards.

This research has made use of the Astrophysics Data System, funded by NASA under Cooperative Agreement 80NSSC21M0056.

\section*{Data Availability}

Most of the underlying data used in this work is in the public domain. The rest of the data underlying this article will be shared on reasonable request to the corresponding author.



\bibliographystyle{mnras}
\bibliography{HephaistosIII} 







\bsp	
\label{lastpage}
\end{document}